\documentclass[sigplan,10pt]{acmart}
\renewcommand\footnotetextcopyrightpermission[1]{}
\usepackage{xspace}
\usepackage{tikz}
\usepackage{color, colortbl}
\usepackage{listings, listings-rust}
\usepackage{booktabs}
\usepackage{subcaption}

\usepackage{listings}

\newcommand{\flock}{Flock\xspace}

\definecolor{lblue}{RGB}{191, 207, 255}
\definecolor{lred}{RGB}{255, 165, 165}
\definecolor{lgray}{RGB}{230, 230, 230}

\definecolor{dkgreen}{rgb}{0,0.6,0}
\definecolor{gray}{rgb}{0.5,0.5,0.5}
\definecolor{mauve}{rgb}{0.58,0,0.82}

\newcommand*\circledone{\tikz[baseline=(char.base)]{
            \definecolor{c1}{RGB}{255, 165, 165}
            \node[shape=circle,fill=c1,inner sep=0.7pt] (char) {\textcolor{black}{\small{1}}};}}
\newcommand*\circledtwo{\tikz[baseline=(char.base)]{
            \definecolor{c1}{RGB}{220, 237, 194}
            \node[shape=circle,fill=c1,inner sep=0.7pt] (char) {\textcolor{black}{\small{2}}};}}
\newcommand*\circledthree{\tikz[baseline=(char.base)]{
            \definecolor{c1}{RGB}{191, 207, 255}
            \node[shape=circle,fill=c1,inner sep=0.7pt] (char) {\textcolor{black}{\small{3}}};}}
\newcommand*\circledfour{\tikz[baseline=(char.base)]{
            \definecolor{c1}{RGB}{149, 225, 211}
            \node[shape=circle,fill=c1,inner sep=0.7pt] (char) {\textcolor{black}{\small{4}}};}}

\begin{document}
\title{Flock: A Low-Cost Streaming Query Engine on FaaS Platforms} 

\author{Gang Liao}
\authornote{Work performed during PhD at University of Maryland, College Park}
\affiliation{%
  \institution{University of Maryland College Park}
}
\email{gangliao@meta.com}

\author{Amol Deshpande}
\affiliation{%
  \institution{University of Maryland College Park}
}
\email{amol@umd.edu}

\author{Daniel J. Abadi}
\affiliation{%
  \institution{University of Maryland College Park}
}
\email{abadi@umd.edu}

\begin{abstract}
Existing serverless data analytics systems rely on external storage services like S3 for data shuffling and communication between cloud functions. While this approach provides the elasticity benefits of serverless computing, it incurs additional latency and cost overheads. We present \flock, a novel cloud-native streaming query engine that leverages the on-demand scalability of FaaS platforms for real-time data analytics. \flock utilizes function invocation payloads for efficient data exchange, eliminating the need for external storage. This not only reduces latency and cost but also simplifies the architecture by removing the requirement for a centralized coordinator. \flock employs a template-based approach to dynamically create cloud functions for each query stage and a function group mechanism for handling data aggregation and shuffling. It supports both SQL and DataFrame APIs, making it easy to use. Our evaluation shows that \flock provides significant performance gains and cost savings compared to existing serverless and serverful streaming systems. It outperforms Apache Flink by 10-20x in cost while achieving similar latency and throughput.
\end{abstract}

\settopmatter{printfolios=true}
\maketitle
\pagestyle{plain}

\lstdefinestyle{base}{language=SQL,
  basicstyle={\small\ttfamily},
  breakatwhitespace=true,
  breaklines=false,
  classoffset=0,
  columns=flexible,
  commentstyle=\color{gray},
  framexleftmargin=0.25em,
  frameshape={}{}{}{}, 
  commentstyle = \color{olive}\bfseries,
  keywordstyle=\color{dkgreen}\bfseries,
  numbers=none, 
  numberstyle=\tiny\color{gray},
  showstringspaces=false,
  stringstyle=\color{mauve},
  moredelim=**[is][\color{dkgreen}\bfseries]{@}{@},
}

\vspace{-3mm}
\section{Introduction}

\renewcommand{\arraystretch}{1}
\begin{table*}
\small
\centering
  \begin{tabular}{lccccccc}
    \toprule
    & \textbf{SQL} & \textbf{SIMD} & \textbf{External Comm. Medium} & \textbf{Hardware} & \textbf{Client Coordinator} &\textbf{Type} & \textbf{Codebase} \\
    \midrule
\textbf{Locus}~\cite{pu2019shuffling} & No & No &  ElastiCache, S3 & x86\_64 & Yes & OLAP & Python \\
\textbf{Lambada}~\cite{muller2020lambada} & No & No &  DynamoDB, SQS, S3 & x86\_64 & Yes & OLAP & Python, C++\\
\textbf{Starling}~\cite{perron2020starling} & No & No &  S3 & x86\_64 & Yes & OLAP & C++ \\
\textbf{Caerus}~\cite{zhang2021caerus} & Yes & No & Jiffy~\cite{khandelwal2022jiffy}, S3 & x86\_64 & Yes & OLAP & Python \\
\textbf{Flock} & \textbf{Yes} & \textbf{Yes} & \textbf{No} & \textbf{arm64}, x86\_64 & \textbf{No} & \textbf{Streaming} & Rust \\
    \bottomrule
  \end{tabular}

\caption{Comparison with Existing Serverless Data Analytics Systems.}
  \vspace{-8mm}
\label{faas_systems}
\end{table*}

High-volume data sources, such as sensor measurements, machine logs, user interactions on websites and mobile applications, typically operate in real time. Stream processing systems play a vital role in delivering the most up-to-date data, empowering organizations to make faster and better-informed automated decisions. These systems must exhibit high performance, elasticity, availability, and ease of use while maintaining cost-effectiveness. Nevertheless, the inherently dynamic and unpredictable nature of streaming workloads~\cite{kulkarni2015twitter,blog2017stream} poses significant challenges in effectively provisioning and configuring resources.

The benefits of cloud computing have spurred recent efforts to migrate streaming analytics applications to fully managed services, such as Google DataFlow~\cite{gcp_dataflow, 43864, akidau2013millwheel, 35650} and AWS Kinesis Data Analytics for Flink~\cite{aws_kinesis_data_analytics}. These Backend as a Service (BaaS) models offer greater elasticity and eliminate upfront costs compared to on-premises alternatives. However, their scaling process can take minutes, making it impractical on a per-query basis.

In contrast, Function as a Service (FaaS)~\cite{awsLambda, gcp_function, azure_functions} fulfill the promise of transparent resource elasticity in the cloud~\cite{jonas2019cloud, hellerstein2018serverless, castro2019rise}. FaaS enables developers to decompose applications into short-lived functions, offering ease of programming, rapid elasticity, and fine-grained pricing. This makes FaaS an appealing solution for streaming processing, as it can accommodate spiky demand through granular resource scaling. FaaS provides more fine-grained elasticity than BaaS, with sub-second start-up times and millisecond-level billing precision. For example, AWS Lambda~\cite{awsLambda} bills customers for the execution time consumed at a 1-millisecond granularity~\cite{awsLambda1ms}, while Kinesis~\cite{aws_kinesis_data_analytics} charges hourly based on the number of Kinesis Processing Units (KPUs) used. This fine-grained billing model makes FaaS cost-effective for low-demand scenarios and allows automatic scaling to handle high loads, with costs proportional to the consumed resources.

We present \flock, a cloud-native streaming query engine that runs on FaaS platforms to explore the potential of function services for stream processing. Table~\ref{faas_systems} highlights the differences between Flock and other state-of-the-art data analytics systems on FaaS platforms~\cite{pu2019shuffling, muller2020lambada, perron2020starling,zhang2021caerus}. While existing approaches leverage the elasticity of cloud object storage services like Amazon S3~\cite{awss3} for data shuffling, this increases performance costs and compromises the advantages of serverless systems. Instead, Flock passes data through invocation payloads between cloud functions, providing a general solution that supports multi-cloud platforms~\cite{awsLambda, gcp_function, azure_functions}. With current limits of 6 MB for synchronous invocations and 256 KB for asynchronous invocations on AWS Lambda~\cite{lambda_quotas}, 32 MB HTTP request size on Google Cloud Functions~\cite{gcp_quotas}, and 100 MB HTTP request length on Azure Functions~\cite{azure_quotas}, Flock can store complete objects directly in the query workflow state, eliminating the need for external storage. Under the FaaS billing model, users pay for each job's duration, proportional to the aggregated runtimes across its component tasks (cloud functions), rather than payload size. This functional programming paradigm reduces latency and execution costs by storing data directly in the workflow. 

Moreover, payload invocation eliminates the need for a query coordinator in the data architecture. As Flock does not rely on external storage for communication between functions, there is no requirement for a coordinator to monitor query stage completion and initiate new stages once dependencies are met. 
Flock employs a unique way for passing multiple payloads/partitions to the same function instance and utilizes shared data structures to ensure exactly-once data aggregation on function services. When checkpointing is enabled, query states are persisted upon checkpoints to prevent data loss and ensure consistent recovery.

Overall, we make the following contributions in this paper:

\noindent $\bullet$ We present Flock, a novel cloud-native streaming query engine that leverages FaaS platforms for real-time analytics. Flock utilizes function invocation payloads for efficient data exchange, eliminating the need for external storage. This not only reduces latency and cost but also simplifies the architecture. It supports both SQL and DataFrame APIs, making it easy to use and allowing users to avoid the time-consuming process of manually translating SQL into cloud workflows.

\noindent $\bullet$ We introduce a template-based approach to dynamically create cloud functions for query stages and a function group mechanism for handling data aggregation and shuffling. 

\noindent $\bullet$ We introduce support for vectorized processing on ARM processors in Flock, delivering a 20\% speedup and reducing costs by more than 30\% compared to x86\_64. 

\noindent $\bullet$ We propose an approach for eliminating the need for a centralized query coordinator by leveraging the DAG representing the relationships between functions, allowing each function to automatically invoke its subsequent child nodes (functions) through direct function invocations. This enables seamless data flow and coordination without the need for a dedicated coordinator, akin to functional programming.

\noindent $\bullet$ We evaluate its performance and cost using the NEXMark and Yahoo Streaming Benchmarks (YSB). Our experiments demonstrate that Flock reduces costs by more than an order of magnitude compared to Flink deployed on EC2 instances, without compromising system throughput or query time.


\vspace{-3mm}
\section{Background}
\label{subsec:streaming}

Stream processing workloads involve continuous data arrivals from diverse sources, necessitating incremental processing. The processing function is unaware of the data stream's start or end points. Consequently, temporal windows~\cite{window_functions} are commonly employed to process this type of data. \flock natively supports tumbling, sliding, and session window functions, enabling users to launch complex stream processing jobs with minimal effort. The initial query stage comprises data source functions that continuously fetch messages from the stream until a complete batch is obtained or the time window expires.



To illustrate the semantics of queries in Flock, consider the following example of a hypothetical online auction system with two tables:%
\begin{lstlisting}[
style=base,
 basicstyle={\scriptsize\ttfamily},
  numbers=left,
   numberstyle=\footnotesize\color{gray},
  xleftmargin=3mm,
]
CREATE TABLE Auction (id INT, item_name VARCHAR(128),
  description VARCHAR(255), initial_bid INT, reserve INT,
  date_time DATE, expires DATE, seller INT, category INT);
  
CREATE TABLE Bid (
  auction INT, bidder INT, price INT, date_time DATE);
\end{lstlisting}%
The \texttt{Auction} table contains all items under auction, and the \texttt{Bid} table contains bids for items under auction.
At some point, the user executes a continuous query to determine the average winning bid price for all auctions in each category across a series of fixed-sized, non-overlapping, 10-second contiguous time periods\footnote{We assume that the auctions are very short-lived (with expiry times less than 10 seconds) and that each auction starts and ends within a single window.}. In Flock, this query is expressed using the following DML:

\begin{lstlisting}[
style=base,
 basicstyle={\scriptsize\ttfamily},
  numbers=left,
   numberstyle=\footnotesize\color{gray},
  xleftmargin=5mm,
  captionpos=b,label=code:query_example
]
-- Flock Context: Window::Tumbling(Schedule::Seconds(10));
SELECT category,
       Avg(final)
FROM   (SELECT Max(price) AS final,
               category
        FROM   auction AS A
               INNER JOIN bid AS B
                       ON A.id = B.auction
        WHERE  B.date_time BETWEEN A.date_time AND A.expires
        GROUP  BY A.id, A.category) AS Q
GROUP  BY category;
\end{lstlisting}

When the user submits this query, Flock continuously and transparently executes it in a microbatch mode on the cloud functions.

\vspace{-2mm}
\section{System Architecture}

\begin{figure}
\centering
\includegraphics[width=0.44\textwidth]{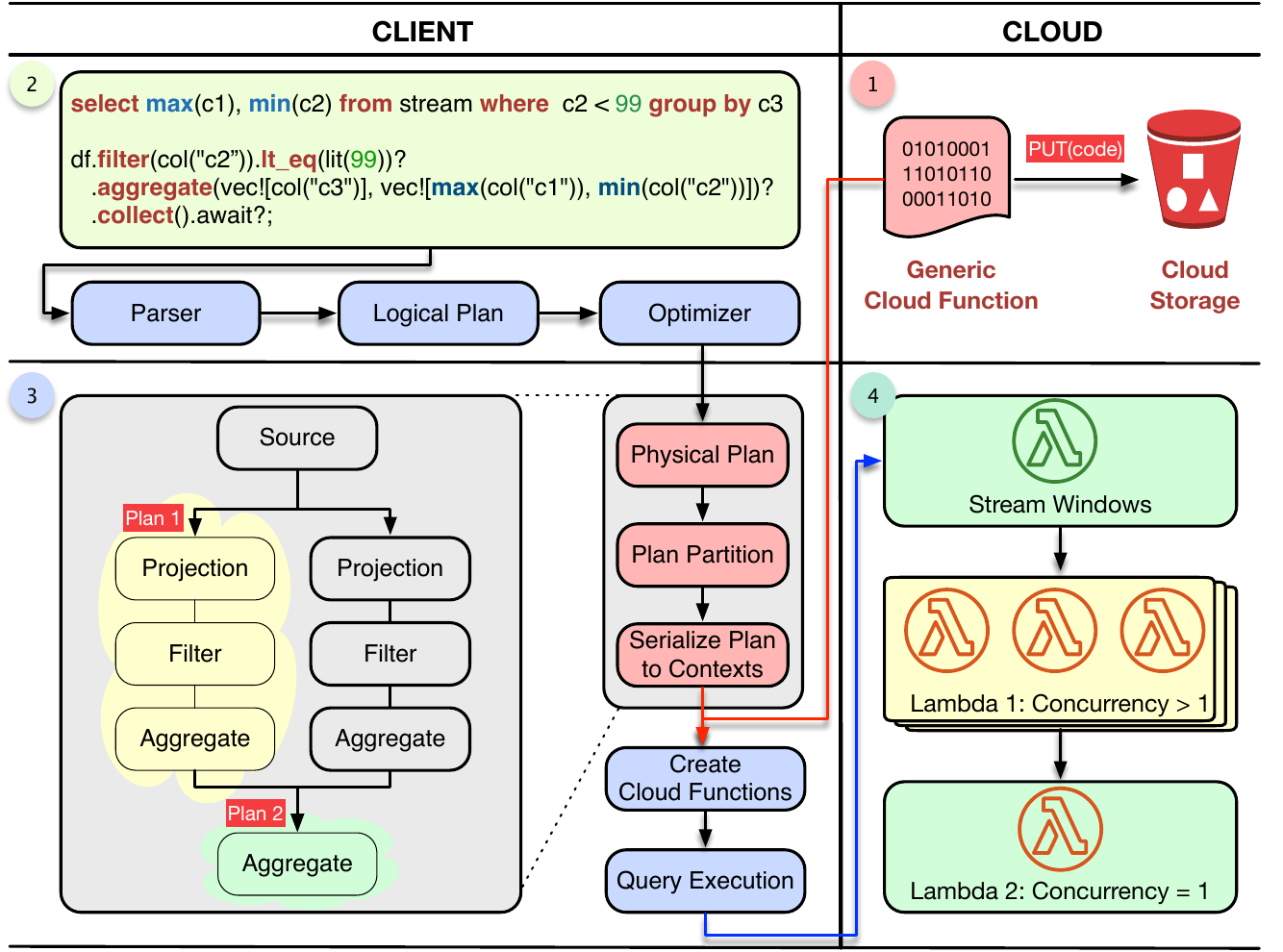}
      \vspace{-4mm}
\caption{System Architecture.}
      \vspace{-5mm}
\label{scq}
\end{figure}

Flock is a cloud-native SQL query engine for event-driven analytics on cloud function services.
Figure~\ref{scq} illustrates the \flock's high-level architectural design. 
\circledone{} The cloud service provider periodically packages and compiles the latest query engine code into a generic cloud function binary, storing it in cloud object storage. \circledtwo{} Upon receiving a SQL query, it undergoes parsing, optimization, and planning into a sequence of low-level operators chosen by the optimizer for efficient execution.
\circledthree{} Flock breaks the execution plan into stages, each comprising a chain of operators with the same partitioning. These stages are serialized as strings and included in the cloud function context. Flock creates cloud functions by retrieving the executable binary from cloud storage and passing the encoded string (cloud context) as a function argument through the cloud vendor's SDK. \circledfour{} Cloud functions are instantiated instantly, enabling real-time query processing. Function arguments are deserialized as the cloud context during initial instantiation, customizing each function for a specific sub-plan. Functions are aware of their role in executing a sub-plan and sending results to the next function, facilitating data flow without a client coordinator.

\subsection{SQL Interface}

While some exploratory research has explored data analytics on cloud services~\cite{pu2019shuffling, perron2020starling, muller2020lambada}, there are no SQL-on-FaaS engines for data analytics yet. End-users currently have to manually split the physical plan for each query when merging query stages into cloud functions as part of a dataflow execution paradigm on the cloud. Requiring users to leverage cloud vendor lock-in APIs to orchestrate query stages is akin to forcing them to create query execution plans directly in database systems. User-generated plans may be suboptimal, leading to significant performance losses, and such customized directed acyclic graphs (DAGs) are error-prone and rarely reusable.
Moreover, some cloud customers have voiced concerns about vendor lock-in, fearing reduced bargaining power when negotiating prices with cloud providers. The resulting switching costs benefit the largest and most established cloud providers, incentivizing them to promote complex proprietary APIs resistant to de facto standardization. Standardized and straightforward abstractions, such as SQL and Dataframe APIs supported by Flock, would remove the most prominent remaining economic hurdle for serverless adoption.

\subsection{Microbatch Execution}


Flock operates in a micro-batch execution mode, similar to Apache Spark's Structured Streaming~\cite{zaharia2013discretized,armbrust2015spark,armbrust2018structured}, which processes data streams as a series of micro-batch tasks, achieving exactly-once fault-tolerance guarantees. In this mode, epochs are typically set to a few hundred milliseconds to a few seconds, with each epoch executing as a traditional analytical job composed of a DAG of functions. Compared to the continuous operator model~\cite{abadi2005design,chandrasekaran2003telegraphcq,golab2003issues}, micro-batch and FaaS are more natural fits for two main reasons: (1) Cloud functions are billed based on the number of invocations and duration, making record-by-record processing orders of magnitude more expensive; and (2) Some cloud providers, e.g., AWS Lambda, only allow a function instance to execute one request at a time, causing dramatic latency increases with a huge number of requests (via record-by-record).

During query planning, Flock automatically chains together sequences of functions, each corresponding to a query stage. Flock implicitly invokes the first cloud function to trigger the execution workflow at recurring times. Although all created functions have the same binary code, when a function is instantiated in the cloud, its environment variable contains the specific cloud context carried during creation, allowing different function instances to be specialized through the context (see Section~\ref{sec:func_template}). Functions share states by passing arguments/payloads and return values to each other, which does not incur additional costs. To send shuffled states to the same function instance without an external communication medium, Flock sets the stateful function's concurrency to one and allocates global memory that allows the function to reuse "static context" across multiple invocations to the same instance. Section~\ref{subsec:func_group} describes how Flock mitigates hotspots in more detail.

\subsection{Fault Tolerance}

\textbf{State Management}. Flock achieves fault tolerance through the employment of a write-ahead log and a state store, both running over an object storage system such as S3 to allow parallel access\footnote{Starting from December 2020, all S3 \texttt{GET}, \texttt{PUT}, and \texttt{LIST} operations are now strongly consistent~\cite{s3strongconsistency}.}. (1) The log keeps track of which data has been processed from each input source and reliably written to the output sink. (2) The state store holds snapshots of operator states for aggregate functions. Similar to Spark Streaming~\cite{armbrust2018structured}, states are written asynchronously and can be behind the latest data written to the output sink. In the event of a failure, the system will automatically track the last updated state in its log and recompute state from that point in the data.


\noindent \textbf{Invocation Failure}. If a cloud function times out or is terminated, the computed end-result remains accurate, with no data loss. This is because the new function is resumed using the most recently stored checkpoint and states from S3. However, unlike traditional nodes, function invocation errors can occur when the invocation request is rejected by issues with request parameters  and resource limits or when the function's code or runtime returns an error. In the case of the asynchronous invocation fails, Lambda retries the function since the payload is part of the invocation, ensuring no data is lost. When an event fails all processing attempts or expires without being processed, it's placed into a dead-letter queue (DLQ)~\cite{lambda-dlq} for further processing, which is part of a function's version-specific configuration.  For synchronous invocation failures, \flock implements a linear backoff algorithm for automatic retries (see Section~\ref{subsec:no_coordinator}).

In asynchronous invocations, the function may receive the same request/payload multiple times because Lambda's internal queue is eventually consistent~\cite{async_invoke}. To avoid double-counting and ensure exactly-once aggregation and processing of each payload, the stateful function maintains a bitmap (see Section~\ref{subsubsec:arena}).

\vspace{-2mm}
\section{Function Templates}
\label{sec:func_template}

\subsection{Template Specialization}
\label{subsec:client-template-spec}
Legal cloud functions are limited to scripts or compiled programs, leading many systems~\cite{perron2020starling} to embed the physical plan into the function code during the code generation phase. These systems generate code for individual tasks, compile it, and package it with necessary dependencies. To execute a job, a scheduler launches tasks as serverless functions and monitors their progress. However, compiling cloud functions and dependencies at query runtime can cause delays of seconds or even minutes, slowing query response time.  For example, Flock is a Rust-based cloud-native query engine. Building an \texttt{x86\_64-unknown-linux-gnu} release version with SIMD and mimalloc/snmalloc features~\cite{mimalloc, snmalloc} on an AWS EC2 instance (c5a.4xlarge) takes roughly 4 minutes, even with incremental compilation. Building from scratch is even more time-consuming, taking 8m 33s due to Flock's lengthy dependency tree. In contrast, Locus~\cite{pu2019shuffling}, built on Pywren~\cite{jonas2017occupy, pywren}, is a pure Python implementation that omits the code-generation and compilation steps, directly taking task code and execution plan as input. While this approach saves time on compilation, it sacrifices performance and cost-efficiency due to longer charged durations.

\begin{figure}
\centering
\includegraphics[width=0.32\textwidth]{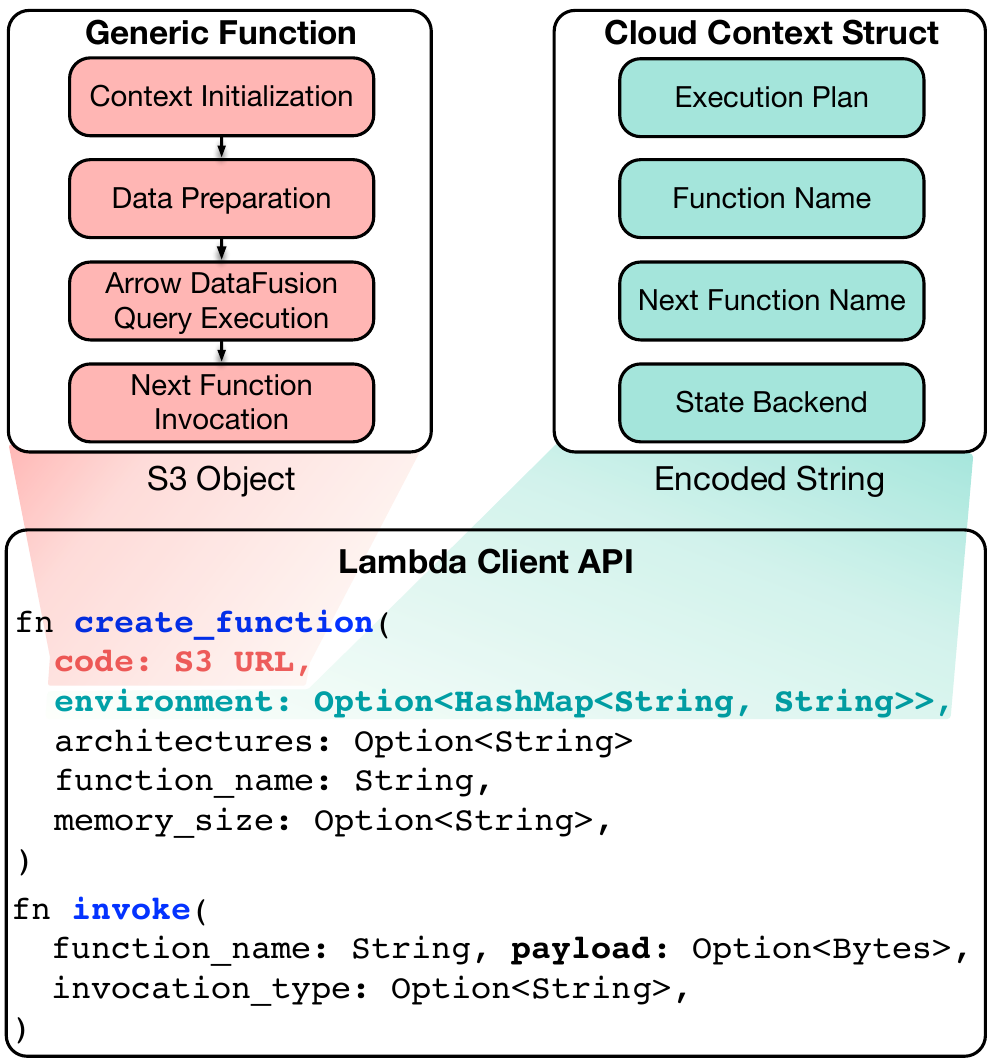}
      \vspace{-3mm}
\caption{Generic Function and Template Specialization.}
      \vspace{-7mm}
\label{template_spec}
\end{figure}

We propose function template specialization as a way to completely eliminate the compilation stage from the query execution pipeline. Template specialization in programming languages allows alternative implementations to be presented based on specific properties of the parameterized type that is being instantiated, enabling certain types of optimization and reducing code bloat. Similarly, as shown in Figure~\ref{template_spec}, Flock's service provider creates, builds, and archives a generic cloud function as a \texttt{bootstrap.zip} file, which consists of four main components: cloud context initialization, data collection and preparation, query execution, and next function invocation (further details are explored in the next subsection). The service provider then uploads the \texttt{bootstrap.zip} to the cloud object storage, and makes a new public release available for Flock users.

Flock eliminates the requirement for the client or central registry to spend time compiling SQL execution plans and new cloud functions into binary code, resulting in significantly lower end-to-end latency\footnote{For developers who want to write custom stream processing logic, Flock's stateful operators are UDFs with state that still require users to compile function code during query runtime. In this case, JIT code generation for each query over LLVM~\cite{lattner2004llvm} or Cranelift~\cite{cranelift} is a better solution to reduce branching overhead and memory footprint.}. Flock creates functions immediately using the S3 object of the generic function that the service provider has provided in advance~\cite{create_func}. Additionally, the cloud context, which includes the execution plan, is serialized and sent as a string into the Lambda API \texttt{create\_function()}'s parameter \texttt{environment} (Figure~\ref{template_spec}). The cloud context is compressed with Zstd~\cite{zstd} after serialization by default, since Lambda environment variables have a default 4 KB service quota that cannot be raised~\cite{lambda_env}. If the execution plan exceeds this limit, Flock stores it in S3 and preserves the S3 object key in the cloud context. The environment variable settings are accessible from the function code during execution on the cloud. This approach reduces query launch time by \textbf{10,000 times}, as launching a cloud function only requires the creation of a function without compilation.

Flock then invokes the newly created function name to execute the query on the cloud function services via the Lambda API \texttt{invoke()}~\cite{invoke_func}. The context initialization is performed once per function instance to prepare the cloud environment for invocations; it reads the encoded string from the environment variable and deserializes it as the cloud context. This process achieves the generic function template specialization. Even though all functions have the same code (i.e., the generic template), each function can identify the specific execution plan and the function to deliver the output to via the cloud context when it is instantiated in the cloud.

\vspace{-2mm}
\subsection{Generic Function}
\label{subsec:generic-func}Flock is a new generation of cloud-native query engine that consists of generic functions and a client library. The generic function can work with any type, rather than a specific type only, allowing it to be designed, built, and delivered to the cloud platform ahead of time. To provide users with the latest query engine capabilities, the cloud service provider only needs to offer an updated version of the generic function on a regular basis without disclosing the source code. The client library can translate SQL queries to executable cloud functions.

A generic function is a function code whose behavior depends on the identities of the arguments supplied to it via \texttt{environment} (see Figure~\ref{template_spec}). When a function is invoked, it deserializes the cloud context provided by the client to discover the appropriate code regions --- those with specializers that are compatible with the actual context. The pseudo code in   Listing\ref{code:generic_func} shows the pseudo-code for how the generic cloud function is implemented and operated. The function code can be broken down into four parts:

(1) \emph{Cloud Context Initialization.}  \texttt{INIT} (line 5) is a synchronization primitive for running a one-time global initialization. The given closure \texttt{ctx\_fn} (line 9-14) is used to deserialize environment variables into cloud context, and it will be run if \texttt{call\_once} (line 15) is used for the first time; otherwise, the routine will not be invoked. Private data that is only used per invocation should be defined within the handler. Global variables such as \texttt{CLOUD\_CONTEXT} retain their value between invocations in the same execution environment. As a result, the cloud context is only initialized once throughout the lifetime of the instance, and future invocations reuse the resolved static context. \texttt{arena} (line 11 and 23) is a type of global resource that are created during initialization stays in memory between invocations, allowing the handler to collect states across invocations.  We explain it in more details in Section~\ref{subsubsec:arena}.

(2) \emph{Data Preparation.}
The function essentially receives the payload in JSON format from the HTTP request's body, computes the result, and either returns it to the client or forwards it to the next functions as HTTP requests.  When the runtime receives an event (line 22), it passes the event to the function handler. \flock leverages Apache Arrow~\cite{arrow-rs} to save streaming data (line 24) in the in-memory columnar format to maximize cache locality, pipelining and SIMD instructions on modern CPUs. 
In the case of the function associated with the aggregate operation, such as \texttt{HashAggregateExec}, \flock uses \texttt{Arena} to collect all data partitions before being given to the embedded query engine in the current function.  More details are described in Section~\ref{subsubsec:arena}.
 

(3) \emph{Query Execution.}
The function includes Arrow DataFusion~\cite{arrow-datafusion}, an in-memory query engine that provides both a DataFrame and SQL API for querying CSV, Parquet, and in-memory data. DataFusion leverages the Arrow~\cite{arrow-rs} compute kernels for vectorized query processing. All rows with a particular grouping key are in the same partitions, such as the case with hash repartitioning on the group keys. Data partitions are processed in parallel in the cloud function (line 26).

(4) \emph{Next Function Invocations.}
Following the execution of the query stage in the current function, the output is placed into the next function invocation's payload (see \texttt{invoke()} in Figure~\ref{template_spec}), and finally, a synchronous or asynchronous invocation (line 27) is made to make distributed dataflow possible. The implicit invocation chain is analogous to functional programming. More complex data shuffling are described in detail in Section~\ref{subsubsec:shuffling}.
\lstset{language=Rust,
  breaklines=true,
  style=boxed,
   basicstyle={\scriptsize\ttfamily},
  frame=none,
  backgroundcolor=\color{white},
    numberstyle=\footnotesize\color{gray},
}
\begin{lstlisting}[caption={Generic Function Skeleton.},captionpos=b,label=code:generic_func]
use lambda_runtime::{service_fn, LambdaEvent};
use serde_json::Value;

/// Initialize the function instance once and only once.
static INIT: Once = Once::new();
static mut CLOUD_CONTEXT = CloudContext::Uninitialized;

macro_rules! init_cloud_context {
    let ctx_fn = || match std::env::var(&**CONTEXT_NAME) {
        Ok(s) => { CLOUD_CONTEXT = CloudContext::Lambda((
            ExecutionContext::unmarshal(&s), Arena::new()));
        }
        ...
    };
    INIT.call_once(ctx_fn);
    match &mut CLOUD_CONTEXT {
        CloudContext::Lambda((ctx, arena)) => (ctx, arena),
        CloudContext::Uninitialized => panic!("uninitialized!"),
    }
}

async fn handler(event: LambdaEvent<Payload>) -> Result<Value> {
    let (mut ctx, mut arena) = init_cloud_context!();
    let (input, status) = prepare_data(ctx, arena, event)?;
    if status == HashAggregateStatus::Ready {
        let output = collect(ctx, input).await?;
        invoke_next_functions(ctx, output, ...).await
    } else if status == HashAggregateStatus::NotReady {
        Ok(json!("response": "data is not yet ready"))
    } else if status == HashAggregateStatus::Processed {
        Ok(json!("response": "data has been processed"))
    }
}

#[tokio::main]
async fn main() -> Result<()> {
    lambda_runtime::run(handler_fn(handler)).await?;
    Ok(())
}
\end{lstlisting}

\vspace{-5mm}
\subsection{Heterogeneous Hardware} AWS Lambda functions running on Graviton2~\cite{aws_graviton}, an Arm-based processor architecture designed by AWS, deliver up to 34\% better price performance compared to functions running on x64 processors for serverless applications~\cite{lambda_arm}, including real-time data analytics. To provide users with better price-performance, Flock offers function binaries for both x86 and Arm architectures. Users can select different generic function binaries from the AWS S3 bucket to create Lambda functions that operate on x86 and/or Arm processors. Currently, Flock has 4 versions on S3: \texttt{x86\_64-gnu}, \texttt{x86\_64-musl}, \texttt{aarch64-gnu}, and \texttt{aarch64-musl}.

For Lambda functions using Arm/Graviton2 processors, duration charges are 20\% lower than the current pricing for x64. However, the reported performance difference (19\%) between x64 and Arm by AWS may not include SIMD optimization. It remains an open question which architecture performs better on query operations when AVX2 and Arm Neon intrinsics are employed. The Graviton2 processor also supports the Armv8.2 instruction set, which includes the large-system extensions (LSE) introduced in Armv8.1. LSE provides low-cost atomic operations and improves system throughput for CPU-to-CPU communication, locks, and mutexes. We compare the latency and duration cost between the two architectures in Section~\ref{sec:expriment}.

\begin{figure}
\centering
\includegraphics[width=0.41\textwidth]{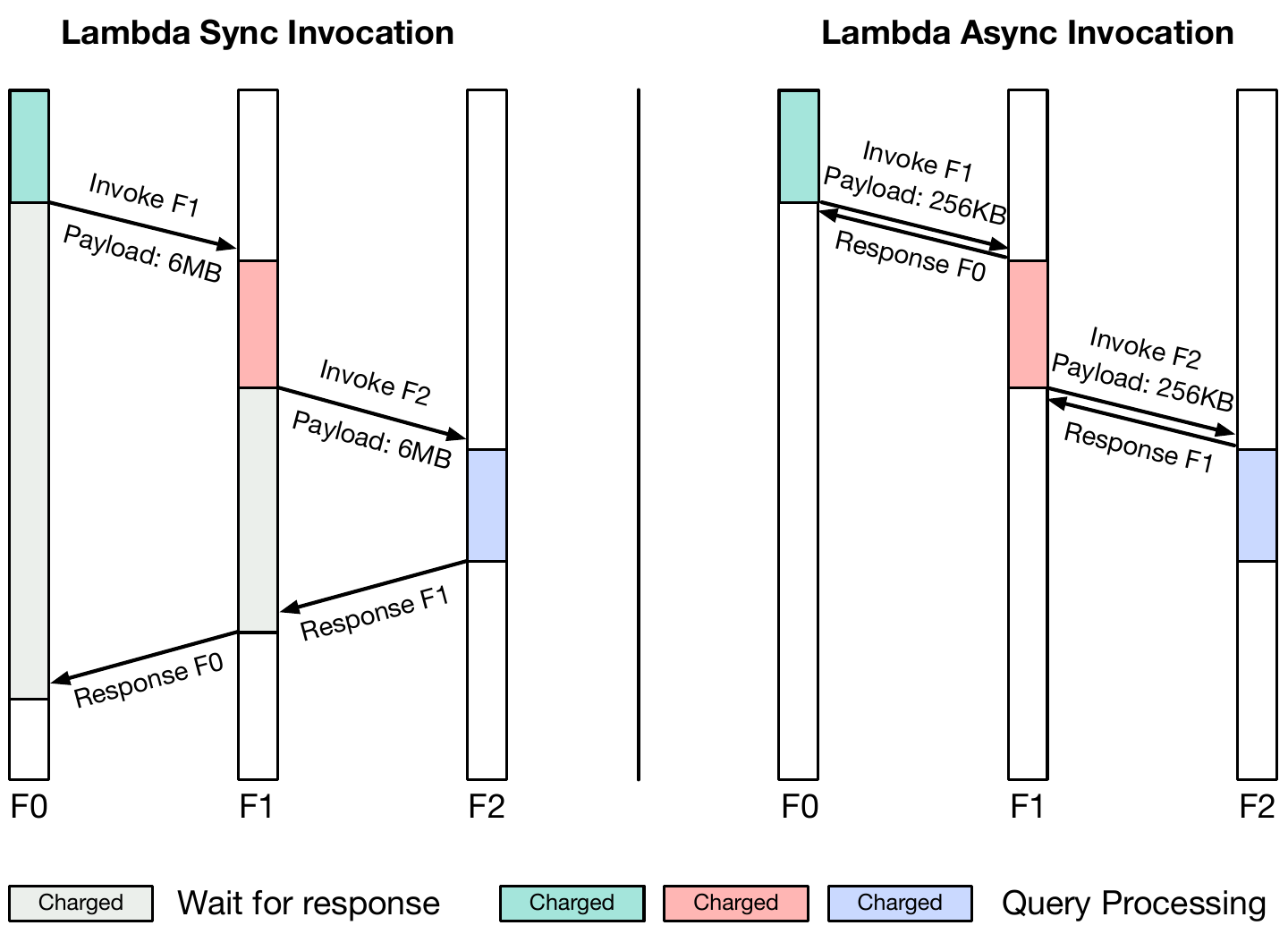}
  \vspace{-4mm}
\caption{The duration charge comparison.}
  \vspace{-5mm}
\label{async_vs_sync}
\end{figure}

 \vspace{-2mm}
\section{Function Communication}
\label{sec:serverless_dataflow}


\subsection{One-Way Road}

Prior work has proposed solutions for data exchange in serverless environments~\cite{klimovic2018understanding, klimovic2018pocket, pu2019shuffling, perron2020starling, muller2020lambada}. These solutions rely on external storage for exchanging large data volumes since cloud functions cannot accept incoming connections. For example, Starling~\cite{perron2020starling} utilizes Amazon S3 to transfer intermediate data between function invocations. However, such solutions introduce additional services, increasing latency (I/O), billable function duration, and S3 access costs, thereby compromising the serverless system's advantages.

In contrast to earlier systems focused on OLAP workloads, \flock is designed for real-time stream processing of gigabyte-scale data. AWS Lambda functions have a 6 MB payload size limit for synchronous invocations and a 256 KB limit for asynchronous invocations~\cite{lambda_quotas}. Function concurrency, the number of instances serving requests simultaneously, has a default regional limit of 1000~\cite{invoke_scaling}, which can be easily increased to 5000 by contacting Amazon. By leveraging these AWS Lambda quotas along with data encoding and compression, \flock can transfer gigabyte-scale intermediate results between functions without relying on external storage.

\flock passes data in the payload when invoking a function (Figure~\ref{template_spec}), serialized as JSON bytes per AWS Lambda's \texttt{application/json} HTTP request body requirement. Data partitioning ensures each partition fits within the payload, enabling seamless data transfer between query stages (functions) without persisting to data stores like DynamoDB~\cite{dynamodb} and S3~\cite{awss3}, incurring no additional invocation costs and reducing billable duration.

Table~\ref{payload_comm} compares the latency of AWS S3 and function payload communication. Objects are compressed using Zstd~\cite{zstd}, achieving a 4x compression ratio on NYC Citi Bike trip data~\cite{citi_bike}, allowing tested partition sizes up to 60 MB (\texttt{15MB * 4}) suitable for streaming workloads. The reported latency incorporates marshalling/unmarshalling and compression/decompression overheads incurred in the Lambda Rust Runtime~\cite{lambda_rust_runtime}, which account for less than 7\% of the total time. For partitions less than or equal to 1.5 MB, payload communication outperforms S3 by an order of magnitude. Since the 15 MB payload size limit is exceeded at larger sizes, \flock executes the same function instance multiple times synchronously or asynchronously, resulting in 6x and 2x speedups, respectively. Section~\ref{subsec:func_group} elaborates on how multiple payloads are routed to the same running instance, a critical aspect of data shuffling.
\vspace{-2mm}
\renewcommand{\arraystretch}{0.8}
\begin{table}[htbp]
\small
  \centering
  \begin{tabular}{@{} c ccc ccc @{}}
    \toprule
     & \multicolumn{3}{c}{AWS S3} & \multicolumn{2}{c}{Lambda Payload} \\
    \cmidrule(r){2-4} \cmidrule(l){5-6}
    Object & Read & Write & Total & Sync & Async \\
    \midrule
    1.5KB & 0.471 & 0.113 & 0.584 & 0.020 & 0.030 \\
    15KB  & 0.471 & 0.144 & 0.615 & 0.020 & 0.044 \\
    150KB & 0.653 & 0.205 & 0.858 & 0.036 & 0.066 \\
    1.5MB & 1.615 & 0.594 & 2.209 & 0.281 & 0.785 \\
    15MB  & 11.720 & 1.828 & 13.548 & 2.201 & 6.054 \\
    \bottomrule
  \end{tabular}
    \caption{The latency comparison (seconds).}
    \vspace{-8mm}
  \label{payload_comm}
\end{table}



\vspace{-2mm}
\subsection{Sync and Async}

When invoking a function asynchronously, AWS Lambda enqueues the event in a Lambda-owned queue and returns immediately, without exposing Lambda's internal queues directly. A separate process dequeues and executes the function. As a multitenant system, AWS Lambda implements fairness through per-customer rate-based limits, with some flexibility for bursting~\cite{lambda_queue_backlogs}. However, occasional invocation delays may occur under heavy workloads.

Figure~\ref{async_vs_sync} illustrates the benefit of asynchronous calling: when the current function invokes the next function, it can return immediately without waiting for the succeeding function to complete execution, significantly decreasing the billable duration. Let $n$ denote the total number of query stages, and $f_i$ represent the lambda function or function group corresponding to the $i$th query stage. The total asynchronous invocation cost is:
\begin{equation*}
\sum _{i=0}^{n}\lambda(f_i) + \sum _{i=0}^{n}d(f_i) 
\end{equation*}

Let $\lambda(f_i)$ denote the cost of function invocations for the $i$th query stage with specific memory and processor configurations, and let $d(f_i)$ represent the billed duration cost of the $i$th query stage. For synchronous invocation, the duration cost (including waiting time) of the $i$th stage is $\sum_{j=i}^{n} d(f_j)$. The total cost of the billing is:
\begin{equation*}
\sum _{i=0}^{n}\lambda(f_i) + \sum _{i=0}^{n}\sum _{j=i}^{n}d(f_j)
\end{equation*}

While asynchronous invocation offers cost benefits for complex, multi-stage queries, synchronous invocation has its advantages in terms of speed, reliability, and cost-effectiveness for simpler workloads. Synchronous invocation is not affected by internal queue throttling, providing faster and more reliable execution. Furthermore, for queries executed by a single function or with shallow stages, synchronous invocation may incur lower billable duration and cost due to its larger 6 MB payload size limit (24 times larger than the 256 KB limit for asynchronous invocations). However, for complex queries with numerous stages, the asynchronous approach may be more cost-effective despite the overhead of additional function invocations, as each invocation is charged for its duration.

\subsection{"No" Coordinator}
\label{subsec:no_coordinator}

Migrating streaming applications from a traditional serverful deployment to a serverless platform presents unique opportunities. In traditional serverful deployments, workflow management frameworks such as MapReduce~\cite{condie2010mapreduce}, Apache Spark~\cite{armbrust2015spark, armbrust2018structured}, Sparrow~\cite{ousterhout2013sparrow}, Apache Flink~\cite{carbone2015apache} provide a logically centralized scheduler for managing task assignments and resource allocation. The scheduler traditionally has various objectives, including load balancing, maximizing cluster utilization, ensuring task fairness, keeping track of distributed tasks, deciding when to schedule the next task (or set of tasks), and reacting to finished tasks or execution failures. Serverless computing does not require a traditional serverful scheduler because FaaS providers are responsible for managing the containers or MicroVMs~\cite{agache2020firecracker} and serverless platforms typically provide a nearly unbounded amount of ephemeral resources. However, existing data systems on FaaS platforms like ~\cite{perron2020starling} and Lambada~\cite{muller2020lambada} still require a coordinator to monitor task completion and start new stages once dependencies are completed due to their use of S3 as the communication medium between functions. Without a coordinator, these systems would have no way of knowing if the current query step is complete.

\flock eliminates the coordinator by putting the name of the next function stage in the current function's cloud context during client-side query planning (see Figure~\ref{template_spec}). When the current function finishes computation, it passes the result to the next function invocation's payload. For asynchronous invocation, if the function terminates abnormally or throws invocation errors, AWS Lambda retries the function. \flock configures a dead-letter queue~\cite{lambda-dlq} on the function to capture events that weren't successfully processed for further processing.
For synchronous invocation, \flock implements a truncated linear backoff algorithm that uses progressively longer waits between retries for rate limit exceeded errors. These retries are only required when \flock passes multiple payloads to a single function with concurrency equal to 1 (see Section~\ref{subsec:func_group}). The current function regularly re-invokes a failed function, increasing the wait time between retries until reaching the maximum backoff time. The wait duration is calculated as: \texttt{min(50 * increase\_factor + random\_ms, max\_backoff)} where \texttt{increase\_factor} starts at 1 and resets when \texttt{50 * increase\_factor} exceeds \texttt{max\_backoff}. \texttt{random\_ms} is bounded to 100ms to avoid synchronized retry waves. Removing the query engine's core coordinator simplifies coding, operation, and maintenance while potentially reducing query processing time.






\subsection{Function Name}
\label{subsec:func_name}

The cloud function name consists of three parts:
\texttt{\footnotesize Function Name: <Query Code>-<Query Stage ID>-<Group Member ID>}
where the query code is the hash digest of a query, the query stage ID is a 2-digit number representing the position of a stage in the DAG, and the group member ID is the position of the function within its group. The function name does not include a timestamp, allowing the created function to be reused by continuous queries without incurring a cold start penalty. This naming convention ensures that each cloud function is appropriately identified and categorized into a distinct query, enabling \flock to efficiently detect and resolve issues.

\subsection{Function Group}
\label{subsec:func_group}


The cloud function concurrency is the number of instances or execution environments that serve requests at a given time~\cite{lambda_instance}. \flock uses \texttt{put\_function\_concurrency()}~\cite{concurrency_func} to set the maximum concurrency for each function in the query DAG, ensuring that the function can scale on its own while preventing it from growing beyond that point. \flock sets the default concurrency to 1000 for stateless functions (e.g., scan, filter, and projection). Each stateless function is preferentially executed on a data partition containing the same keys to maximize data parallelism. However, if the concurrency of a stateful function (e.g., group by, sort, and join) is set to more than 1, \flock cannot ensure the integrity of the query results. In this case, Lambda is likely to spawn multiple running instances to handle payloads from the non-aggregate functions, causing the partial results to diverge and ultimately fail to aggregate. Therefore, for aggregate functions, \flock sets the concurrency to 1, which enforces AWS Lambda to create at most one running instance for the aggregation function at any given time.

Although one of the key advantages of a serverless query engine is its ability to effortlessly scale to accommodate fluctuating traffic demands or requests, with minimal capacity planning requirements, setting the concurrency of aggregate functions to one contradicts the essence of serverless computing. Since there is only one function instance for the current query stage, and AWS Lambda does not yet support per-instance concurrency~\cite{gcp_concurrency} akin to GCP Functions, which allows concurrent requests on a single running instance, hotspots arise due to the need to await the completion of the preceding aggregate task.

We propose the \emph{function group} technique, which creates a set of cloud functions in a group for each query stage after partitioning the physical plan, in order to mitigate the hotspot effect. (1) For non-aggregate functions, the concurrency is set to 1000 by default. \flock creates only one function member (name) in that group, and AWS Lambda governs the running instances and routes requests to them. (2) For aggregate functions, the concurrency is set to 1. \flock creates a group consisting of multiple identical functions with different names. 

\begin{figure}
\centering
\includegraphics[width=0.4\textwidth]{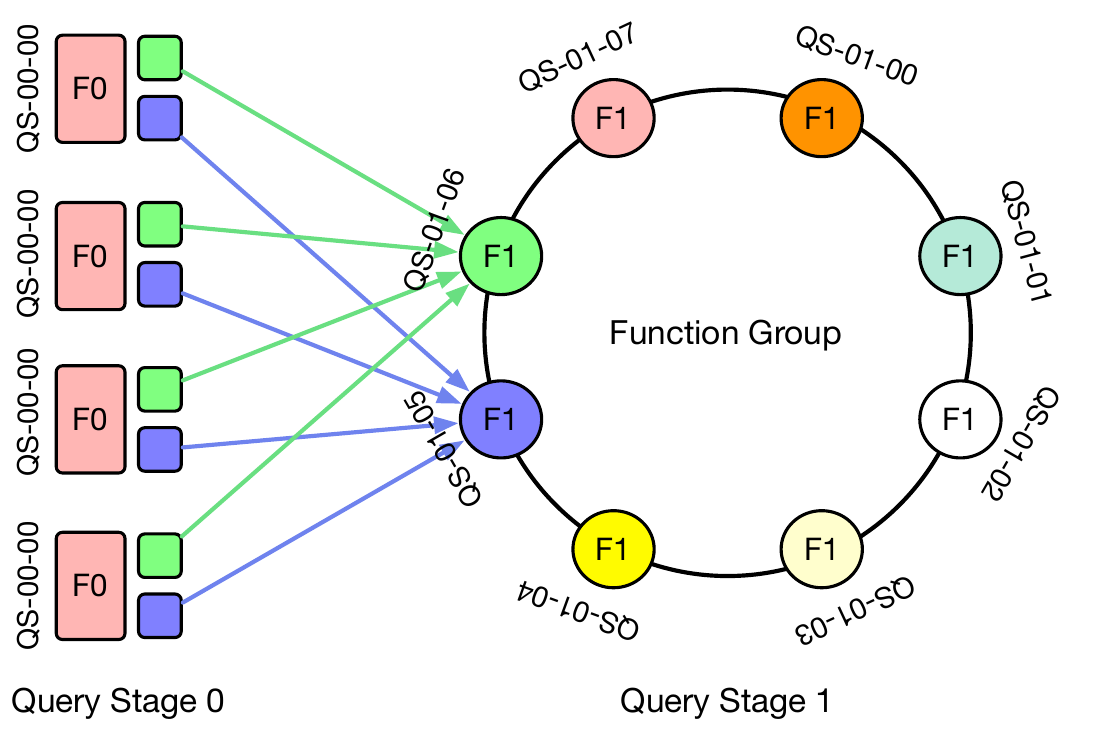}
  \vspace{-4mm}
\caption{Cloud Function Group.}
  \vspace{-6.5mm}
\label{cloud_function_group}
\end{figure}

Figure~\ref{cloud_function_group} illustrates the cloud function group mechanism. For query stage 0, with the default concurrency of 1000, Lambda spawns four instances that perform local aggregation and hash partition the output into two payloads (green and blue). Payloads with the same color across instances share a shuffle ID, used to generate a consistent hash key~\cite{stoica2003chord}. The cloud context specifies the next group as \texttt{Group(QS-01,8)}, where \texttt{QS-01} is the group name, and 8 is the group size. \flock maps payloads with the same key across instances to the same function name in the next stage, distributing shuffle operations while guaranteeing data integrity with a concurrency of 1. Each instance performs a consistent hash lookup to map different partitions to distinct functions in a counterclockwise order, minimizing serial aggregation due to hash collisions. The consistent hashing function is independent of the group size, enabling dynamic scaling by adding or removing functions to balance hotspots and cold starts. Moreover, function instances can dynamically coalesce shuffle partitions for adaptive query execution~\cite{spark_aqe} by reading statistics from the state store.

\section{Dynamic Serving Paradigm}
\label{sec:dataflow_paradigm}

Unlike traditional distributed execution engines, \flock's execution plan is a \textbf{dynamic} directed acyclic graph that evolves over time in the cloud, contrasting with static on-premises plans. This dynamic nature arises because each stage in the query DAG corresponds to a cloud function group, where AWS Lambda automatically scales running instances up or down based on the incoming event volume. However, since data shuffling occurs via function invocation payloads, the aggregate function is likely to receive data partitions from multiple temporal windows. This section addresses the following questions: 1) When data partitions from different shuffling operations or queries are delivered to the same function instance, how can aggregation be distinguished between data? 2) How is the completion of aggregation determined, signaling the transition to the next stage?

\subsection{Payload Structure}
\label{subsubsec:payload}

The payload encompasses a \texttt{Data} field containing an "on-the-wire" representation of Arrow record batches. The \texttt{Schema} field delineates the tables, fields, relationships, and data types carried. It also incorporates metadata pertaining to the data. For instance, the \texttt{Encoding} field offers various compression options, 
facilitating compressed transmission of data.


\begin{lstlisting}[
  basicstyle={\scriptsize\ttfamily},
  columns=flexible,
  numbers=none, 
  frame=none,
  backgroundcolor=\color{white},
]
Payload: { UUID, EpochID, ShuffleID, Data, Schema, Encoding } 
UUID: { QID, SEQ_NUM, SEQ_LEN }
QID: <Query Code>-<Job ID>-<Query Timestamp>
\end{lstlisting}

To enhance the determinism of shuffling and aggregation, \flock assigns each payload a universally unique identifier (UUID).  The \texttt{QID} component of the UUID is derived from the function name (see Section~\ref{subsec:func_name}), but unlike the function name, it also incorporates the query start timestamp and job ID. This allows for differentiation of payloads originating from distinct queries. The \texttt{EpochID} indicates the specific microbatch from which the current data partition originates. In addition to the \texttt{QID}, the payload's UUID comprises \texttt{SEQ\_NUM} and \texttt{SEQ\_LEN}. \texttt{SEQ\_NUM} is a monotonically increasing number that identifies the uniqueness of the payload within a set of aggregated data, while the \texttt{SEQ\_LEN} field represents the total number of payloads to be aggregated. These two fields enable the aggregate function to determine whether all payloads have been collected for a given job.


In the case of partial aggregation within a function, multiple payloads are produced, each potentially being shuffled to different functions in the next function group (see Section~\ref{subsec:func_group}). The Shuffle ID is used to assign an incremental number to each output payload from the function. Payloads across function instances within the same stage that belong to the same partition range are allocated the same Shuffle ID. This mechanism is primarily employed to distinguish different aggregate tasks of the same query job, as they can all be mapped to the same next function. For example, in Figure~\ref{cloud_function_group}, the green payload is assigned a Shuffle ID of 1, while the blue payload is assigned a Shuffle ID of 2.

\begin{figure*}
\centering
\includegraphics[width=0.9\textwidth]{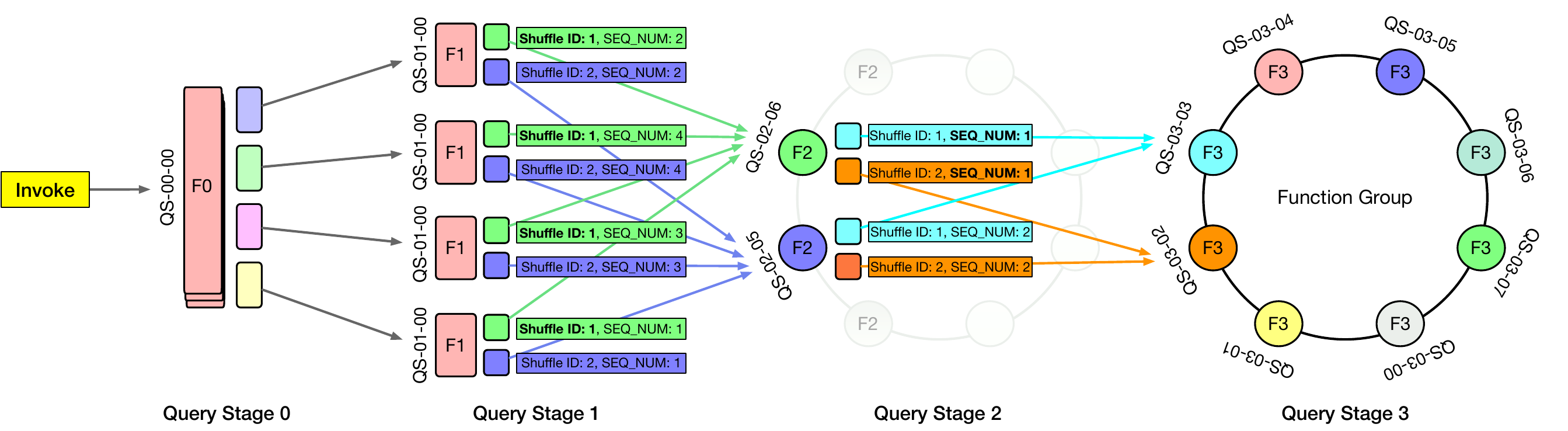}
  \vspace{-4mm}
\caption{Multi-level Shuffling.}
  \vspace{-4mm}
\label{dist_shuffling}
\end{figure*}

\subsection{Global Memory}
\label{subsubsec:arena}

Static initialization occurs before query code execution, allowing functions to reuse global resources across multiple invocations. The \texttt{INIT} code (Listing~\ref{code:generic_func}, line 23) runs when a new execution environment launches or scales up, deserializing the cloud context and creating the memory arena once per environment to avoid redundant loading. The \texttt{Arena} is a global versioned hash map that aggregates data partitions outside the function handler, ensuring data integrity. The key is a tuple of the payload's \texttt{QID} and Shuffle ID, while the value contains the received payloads. The \texttt{SEQ\_LEN} field indicates the total payloads to collect per session. Since AWS Lambda lacks per-instance concurrency~\cite{gcp_concurrency}, \flock decompresses and deserializes data only after receiving all payloads, maximizing parallelization.

For asynchronous invocations, \flock employs a bitmap to handle duplicate payloads resulting from Lambda's eventually consistent internal queue~\cite{async_invoke}. The \texttt{SEQ\_NUM} serves as a bitmap index, representing each payload as a single bit to track the aggregation state, ensuring payloads are processed exactly once while utilizing the bitmap.
Even in cases where the function output is empty, payloads containing solely metadata must be forwarded to the next function. This approach ensures that the aggregate function receives the complete set of group-by data.

\subsection{Multi-level Shuffling}
\label{subsubsec:shuffling}

Consider the query execution plan for an online auction system from Section~\ref{subsec:streaming}, which \flock divides into four stages as depicted in Figure~\ref{dist_shuffling}.

\textbf{Stage 0}: This stage reads upstream streaming data from the \texttt{Bid} and \texttt{Auction} sources until the time window is reached, within the same running instance for simplicity. The repartition operator uses a hash of the join key to map N input partitions to M=4 output partitions, distributing data such that records with the same key value reside in the same partition or payload. \flock invokes the next function 4 times to deliver these payloads to the subsequent query stage.

\textbf{Stage 1}: Lambda spawns four instances, one per input payload with distinct \texttt{SEQ\_NUM} values from 1 to 4. Each instance performs local hash aggregation, repartitioning into two output payloads (green and blue) after the hash join. Output payloads inherit the input \texttt{SEQ\_NUM}. Shuffle IDs increment across output payloads. Using a deterministic seed, each function performs consistent hashing to map payloads counterclockwise to the next functions in parallel.


\textbf{Stage 2}: Unlike Stage 1, the current function collects multiple input payloads in the global arena (Section~\ref{subsubsec:arena}). Output shuffle IDs follow the same allocation, while input shuffle IDs are set to the output \texttt{SEQ\_NUM} to detect duplicate payloads for the same aggregate job at the next stage.

\textbf{Stage 3}: This stage produces output partitions, with the next function delivering results to downstream services.



\section{Evaluation}
\label{sec:expriment}

Flock is a cloud-native streaming execution engine developed entirely in Rust, comprising \~13,000 lines of code (LoC). The SQL component spans 748 LoC, with query operators sourced from the single-host query engine Arrow DataFusion~\cite{arrow-datafusion}, which we extended to support distributed partitioning of query plans, distributed query processing on FaaS platforms, fault tolerance mechanisms, and orchestration via encapsulation of the AWS Lambda SDK. As Flock executes queries by directly invoking AWS Lambda functions, all of our experiments are conducted on the AWS Lambda platform.




\subsection{Benchmark}

We evaluate Flock's performance using two streaming benchmarks: the Yahoo Streaming Benchmark (YSB)\cite{chintapalli2016benchmarking} and the NEXMark Benchmark\cite{tucker2008nexmark}. YSB is a simple advertisement application that reads JSON events from Kafka and stores a windowed count of relevant events per ad campaign in Redis. The NEXMark Benchmark, an evolution of the XMark benchmark for an online auction house, presents a schema with three concrete tables and a set of queries to execute in a streaming context. NEXMark aims to provide a benchmark that extensively utilizes operators while closely resembling a real-world application based on a well-known problem. The Apache Foundation adopted and extended the original NEXMark benchmark for use in Beam~\cite{beam}, a system designed to provide a general API for various streaming systems. To introduce more dynamism, they reduced the window sizes to ten seconds, in contrast to the minutes and hours specified in the original benchmark. Additionally, they incorporated more queries~\cite{nexmark_bench}, including N1 - N8 from the original NEXMark queries and N0 and N9 - N13 from Apache Beam.
We follow the widely adopted Beam implementation.  

\begin{figure}
 \centering
    \begin{subfigure}[b]{0.44\textwidth}
    \includegraphics[width=\textwidth]{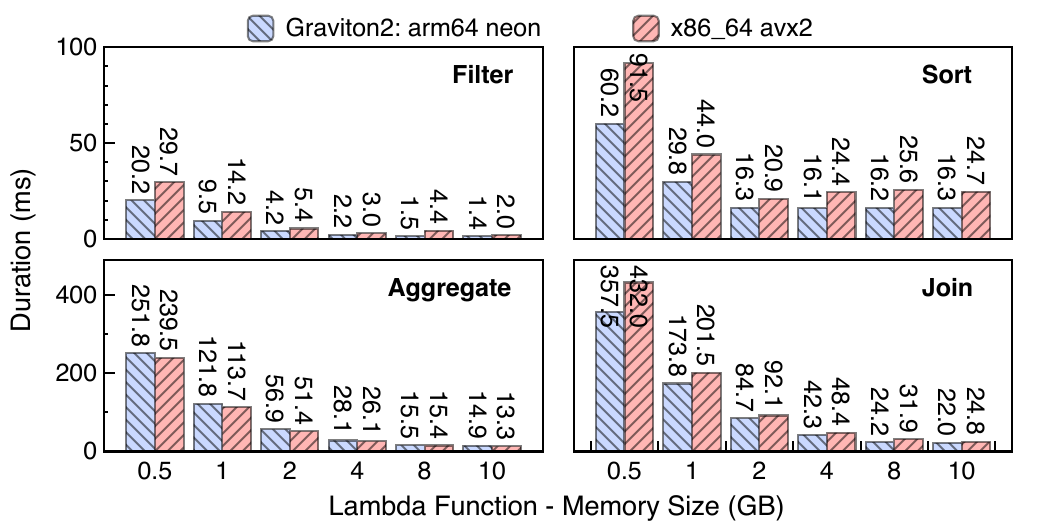}
    \subcaption{Primitive query operators}
    \end{subfigure}
    \vfill
    \begin{subfigure}[b]{0.44\textwidth}
    \includegraphics[width=\textwidth]{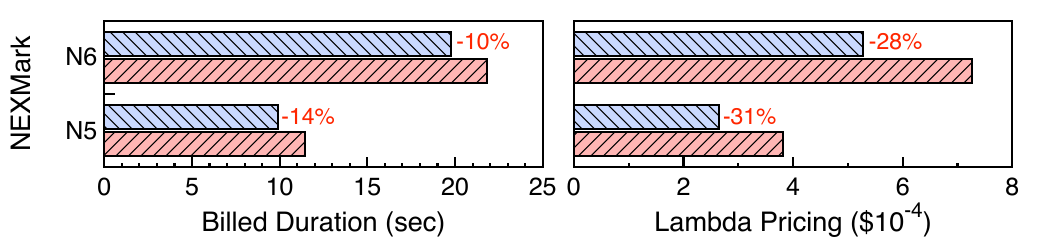}
    \subcaption{NEXMark N5 and N6}
    \end{subfigure}
      \vspace{-4mm}
    \caption{Lambda function on x86 and Arm processors.}
      \vspace{-8mm}
    \label{x86_vs_arm64}
\end{figure}

\begin{figure*}
 \centering
    \begin{subfigure}[b]{0.245\textwidth}
    \includegraphics[width=\textwidth]{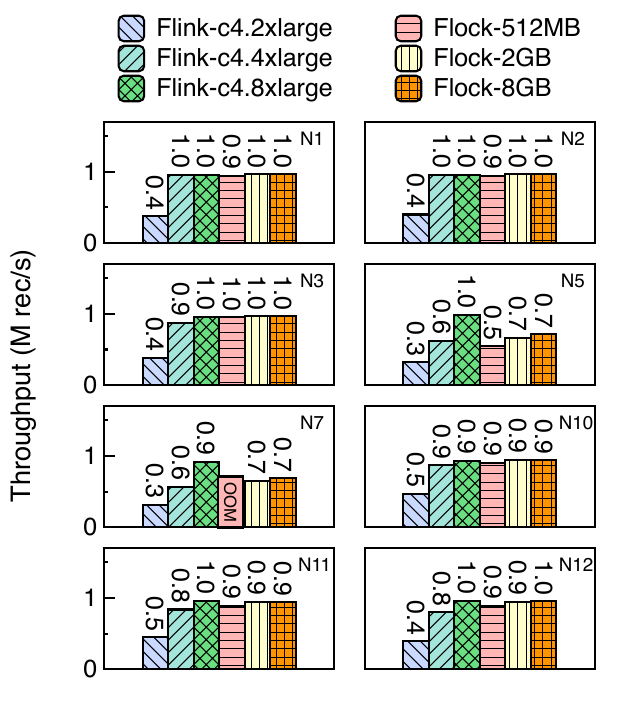}
    \end{subfigure}
    \hfill
    \begin{subfigure}[b]{0.245\textwidth}
    \includegraphics[width=\textwidth]{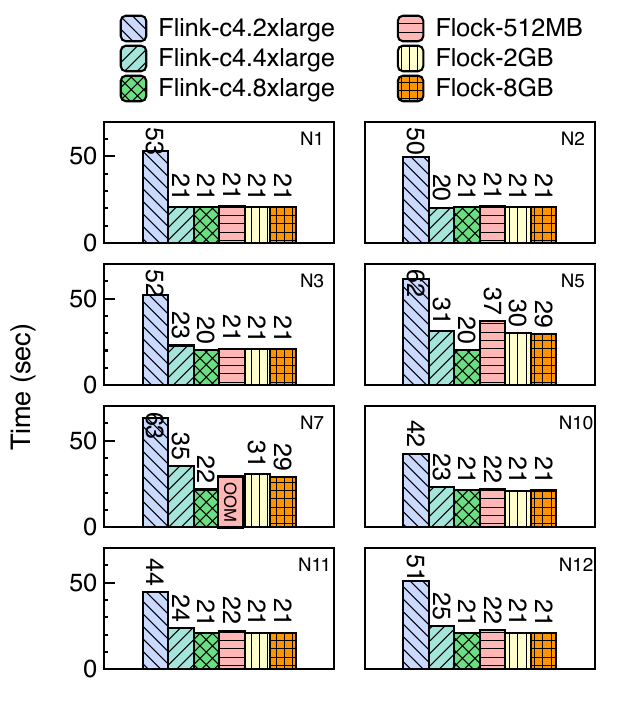}
    \end{subfigure}
    \hfill
    \begin{subfigure}[b]{0.245\textwidth}
    \includegraphics[width=\textwidth]{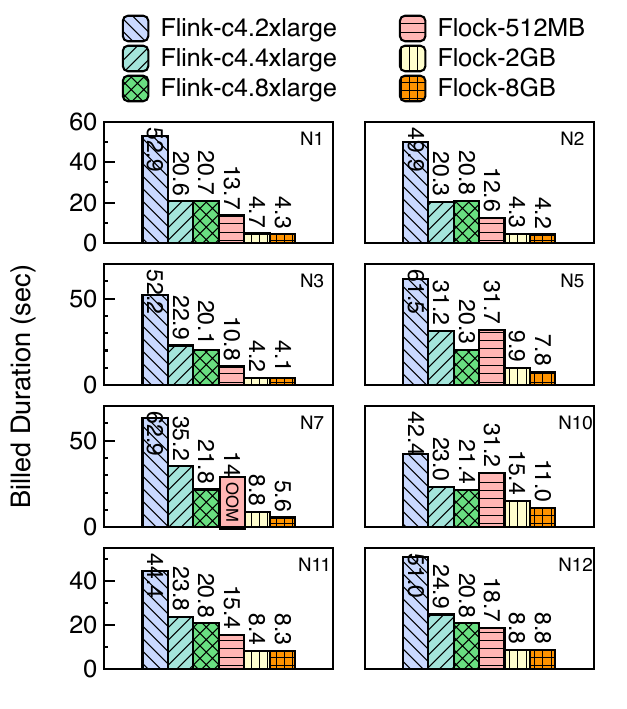}
    \end{subfigure}   
    \hfill
    \begin{subfigure}[b]{0.245\textwidth}
    \includegraphics[width=\textwidth]{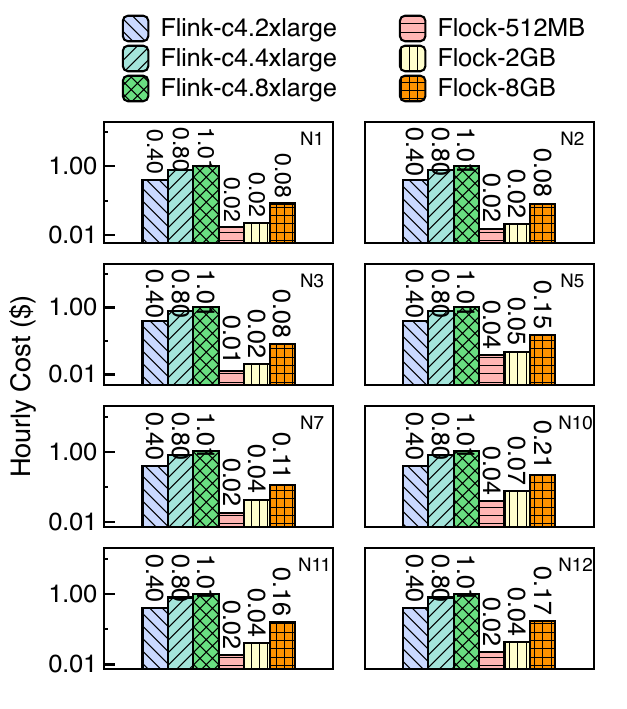}
    \end{subfigure}    
   \vspace{-5mm}
    \caption{Performance cost of executing 20 million NEXMark events and 1 million events per second.}
    \label{nexmark_throughput_latency}
    \vspace{-2mm}
\end{figure*}

 \vspace{-2mm}
\subsection{Heterogeneous Hardware}

This experiment leverages AVX2 and ARM Neon intrinsics, relying on Rust SIMD auto-vectorization and handwritten Arrow kernels that explicitly employ SIMD intrinsics. We generated 500,000 NEXMark events, comprising 9995 person events, 29985 auction events, and 459770 bid events.

Figure~\ref{x86_vs_arm64}(a) illustrates the performance of the Lambda function executing four query operators (filter, join, aggregate, and sort) across x64 and Arm architectures while varying the function's memory size. Each subplot performs a different operation on the events. The proportion of allocated memory determines the dedicated CPU share for a function, necessitating memory tuning to adjust CPU allocation~\cite{lambda-memory}. With the exception of aggregate operations, where Arm is 5-10\% slower than x86, the billed duration for all other operations is lower on Arm. On Arm, the filter's duration accounts for 34\% - 77\% of the total x64 time, the join's duration for 76\% - 91\%, and the sort's duration for 61\% - 76\%. Furthermore, Arm's duration charge is 20\% less expensive per millisecond than x64. For instance, the 1ms charge for ARM 512MB is \$0.0000000067, which is 20\% cheaper than \$0.0000000083 for x64~\cite{lambda-pricing}. Compared to traditional x64 cloud architectures, Flock on the AWS Graviton2 processor achieves greater cost savings due to shorter durations and lower charges.

Figure~\ref{x86_vs_arm64}(b) compares the performance of NEXMark queries N5 and N6 across x64 and Arm architectures. N5 introduces the first usage of windowing in NEXMark, requiring a sliding window that computes hot items over the last 10 seconds and updates every 5 seconds. N6 is the sole query in NEXMark that utilizes the \texttt{partition by} clause. Both queries executed for 20 seconds with an input rate of 1 million events per second. On the Arm architecture, N5 exhibited a 14\% improvement in performance compared to x64, translating to a 31\% cost reduction. Similarly, N6 achieved a 28\% cost savings on Arm. In comparison to x64, the Arm architecture demonstrated superior performance and cost-effectiveness. Consequently, the remaining experiments were conducted on Arm-based Graviton2 processors.

\subsection{Performance Cost}

We evaluated the performance of Flink and \flock using NEXMark queries 1, 2, 3, 5, 7, 10, 11, and 12 in our experiments. N1, N2, and N3 are elementwise queries that feed \flock a micro-batch of events per second. N5 is a sliding window query that schedules overlapping events occurring in the last 10 seconds and updates every 5 seconds. N7 is a tumbling window query that aggregates events using distinct time-based windows opening and closing at 10 second intervals. N10 is a query to log all occurrences to the file system, with Flink saving output to the local file system and \flock saving data to S3. N11 is a session window query that groups events for the same user occurring at similar times while filtering out periods with no available data. N12 is a tumbling window query with a 10 second interval dependent on processing time.

Figure~\ref{nexmark_throughput_latency} compares the throughput, query time, billed duration and hourly cost of executing 10 million events and 1 million events per second between Flink and \flock under different configurations. We deployed Flink on c4.2xlarge, c4.4xlarge and c4.8xlarge EC2 instances. EC2 instances are long-running, we set the duration of Flink to be the same as the query time. In contrast, for \flock, the billed duration specifically refers to the execution part of the cloud functions and excludes data preparation and transmission. We configured the memory sizes for Flock's cloud functions to be 512 MB, 2 GB and 8 GB. To ensure a fair comparison, we used 8 workers for Flink, equal to the concurrency of \flock functions. \flock asynchronously updates the state to S3, while Flink updates the state to the local RocksDB~\cite{rocksdb}. To avoid compaction overhead on the EC2 instances, we only enabled the hashmap state backend for both \flock and Flink in our experiments.

The c4.2xlarge instance has 8 vCPUs and 15.0 GiB of memory, yet its performance is still far inferior to Flock-512MB. This discrepancy can be attributed to Flink's Scala-based implementation, while Flock is a Rust-based high-performance query engine that incorporates SIMD and mimalloc~\cite{mimalloc} and builds upon Arrow DataFusion~\cite{arrow-datafusion}. When using c4.4xlarge or c4.8xlarge instances, Flink generally achieves similar throughput and query time as \flock. For Flock-512MB on N10, the duration exceeds the query time because the processing of events from distinct mini-batches or windows can be separated, allowing the invocation of new cloud functions to process any stacking events. This pipeline parallelism partially hides the duration delay, thereby reducing the query response time. The query time does not fall below 20 seconds since we produce a total of 20 million events and only process 1 million events per second. When running N7 on Flock-512MB, an out of memory error is thrown due to N7 requiring 676 MB of RAM for collecting, decompressing, and deserializing data. Even if the function does not complete successfully, it is still charged for the execution time.


As illustrated in the hourly cost subgraph, Flock can reduce the hourly cost to 1/10 while maintaining similar performance to Flink. When the streaming data rate is low, the volume is modest, or the data is queried infrequently, Flock's cost performance surpasses Flink by more than two orders of magnitude.


\vspace{-2mm}
\subsection{Invocation Payload}
%


\begin{figure}
 \centering
    \includegraphics[width=0.49\textwidth]{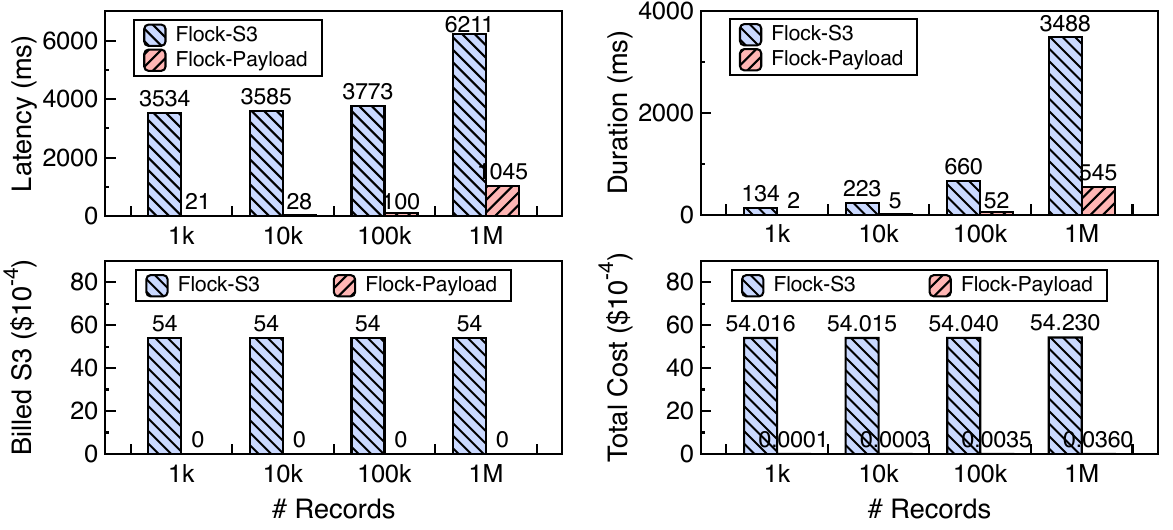}
          \vspace{-7mm}
    \caption{Flock Payload versus Flock S3 on NEXMark Q3.}

    \label{payload_s3_cost}
      \vspace{-6mm}
\end{figure}

Table~\ref{payload_comm} presents the difference in latency between payload and S3 communication when the Lambda memory size is set to 128MB, focusing on the communication aspect rather than end-to-end query processing. It is important to note that the coordinator overhead of state-of-the-art systems, such as Starling~\cite{perron2020starling}, is not included in this comparison. Therefore, Figure~\ref{payload_s3_cost} compares the invocation with payload to S3 communication in terms of latency, duration, and billed cost while varying the number of events on NEXMark Q3. In this experiment, the memory size of the Lambda function is set to 512MB, and it is launched in the \texttt{us-east-1} region, with Flock-S3's coordinator deployed on the client-side.



For Flock-S3 (akin to Starling), the difference between latency and duration, which is approximately 3 seconds, indicates the overhead of the coordinator and function calls. Flock-S3 exhibits an order of magnitude slower performance compared to Flock-Payload due to the round trips between the coordinator and cloud functions. Furthermore, Flock-S3 incurs a one-order-of-magnitude higher billed duration cost than Flock-Payload. This can be attributed to the fact that all S3 reads and writes occur during function execution, and the I/O latency is included in the billed duration. In contrast, increasing the number of events in Flock-Payload results in a larger payload size, which impacts both network transfer time and execution time, consequently affecting query latency. However, the network transfer time does not influence the billed duration in the case of Flock-Payload.


The billed S3 subgraph illustrates the cost of utilizing S3 as an external communication medium for query processing. According to the S3 pricing model~\cite{s3-pricing}, \texttt{PUT}, \texttt{COPY}, \texttt{POST}, and \texttt{LIST} requests are charged at a rate of \$0.005 per 1,000 requests, while \texttt{GET}, \texttt{SELECT}, and all other requests are charged at a rate of \$0.0004 per 1,000 requests. For S3 reads and writes, the billing formula is \texttt{ceil(requests / 1000) * 0.0054}, where \texttt{requests} represents the number of that specific type of request made during a monthly billing interval within a single S3 region. In this case, since the total number of S3 requests is less than 1,000, the charge is directly \$0.0054. Flock-Payload, on the other hand, does not use S3 for data transmission between functions, resulting in no additional cost. The total cost is presented in the last subgraph, with the integer component originating from S3 communication and the fractional part stemming from the duration cost.

\vspace{-2mm}
\subsection{Distributed Aggregation}



Figure~\ref{scale_out_query} presents the latency and billed duration of NEXMark Query 4 (N4) and YSB under both centralized and distributed execution modes. For NEXMark N4, we generated 10 million events. In the centralized mode, \flock invoked the same function instance 65 times to complete the query due to the payload limit. In the distributed mode, ordinary lambda functions have a default concurrency of 1000, while the aggregate function group consists of 8 members, each with a concurrency of 1. By employing the distributed mode, the latency of N4 is reduced by a factor of 4. However, the billed duration is 10 times higher compared to the centralized mode. This can be attributed to the fact that N4 is divided into four query stages, each of which is invoked multiple times due to shuffling or aggregation. Each function execution contributes to the billable duration.

YSB models a simple ad account environment where ad view events enter the system and those of a certain type are accounted to their associated campaign. \flock is expected to report the total ads for each campaign within a tumbling window of 10 seconds. We generate 1 million events per second. Due to the larger size of ad events (all string types) compared to NEXMark events, we increased the Lambda function capacity to 8GB, as the centralized mode cannot process queries in a 2GB memory environment. The centralized mode exhibits a latency \~38 times higher than the distributed mode. This is attributed to the large ad event size, requiring \flock to invoke the same function instance 540 times to collect 10 seconds of window data and run the query. However, the billed duration in the distributed mode is comparable to that in the centralized mode, indicating the clear benefits of distributed query processing for YSB.
\vspace{-4mm}
\renewcommand{\arraystretch}{0.8}
\begin{table}[htbp]
\small
\centering
  \begin{tabular}{c|cccc}
    \toprule
\textbf{Query} & \textbf{Mode} & \textbf{Memory} & \textbf{Latency} & \textbf{Billed Duration} \\
    \midrule
N4 & centralized & 2G & 17.49s & 6.25s \\[0.1pt]
N4 & distributed & 2G & 4.12s & 59.42s \\
YSB & centralized & 8G & 113.38s & 31.54s \\
YSB & distributed & 8G & 2.95s & 33.53s \\
    \bottomrule
  \end{tabular}
\caption{Distributed query processing.}
\vspace{-8mm}
\label{scale_out_query}

\end{table}




\vspace{-4mm}
\subsection{Cold Start}

\begin{figure}
\centering
\includegraphics[width=0.42\textwidth]{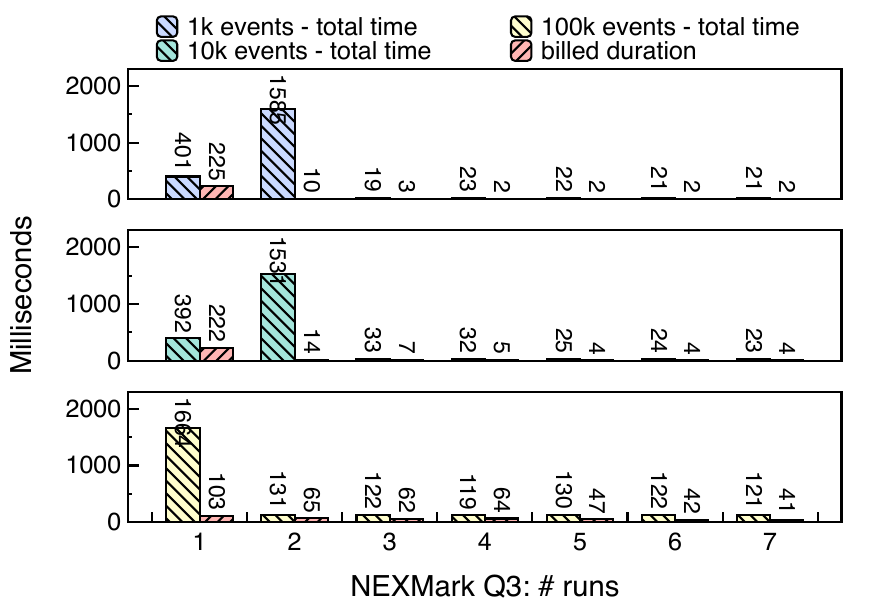}
\vspace{-4mm}
\caption{Lambda cold start cost on NEXMark Q3.}
\label{cold_start_perf}
\vspace{-6mm}
\end{figure}

Figure~\ref{cold_start_perf} illustrates the latency and billed duration of running NEXMark N3 multiple times with varying numbers of events per second. Cold starts, which occur when a new Lambda instance handles its first request, result in longer processing times due to the need for the Lambda service to deploy the code and spin up a new microVM~\cite{agache2020firecracker}. The first request also triggers a one-time function that initializes the Lambda execution context (see line 23 in Listings \ref{code:generic_func}). When the number of events per second is 1K or 10K, both the first and second invocations experience significant delays. The 2nd run's billed duration decreases dramatically, indicating that it is not a new instance, but its overall time increases to 1.6 seconds\footnote{We believe this is some unexplained behavior of AWS Lambda infrastructure.}. 
Warm runs demonstrate a one- to two-order-of-magnitude decrease in latency. For stream processing, as long as the maximum idle time limit is not exceeded, \flock is not impacted by cold runs since Lambda remains warm due to continuous query execution. 


\vspace{-2mm} 
\section{Related Work}

Major cloud providers introduced serverless workflow services\cite{lambda_step_functions, azure_durable_functions, google_cloud_composer}, which provide easier design and orchestration for serverless workflow applications. Netherite~\cite{burckhardt2022msr} and Kappa~\cite{zhang2020kappa} are distributed execution engines that offers high-level language programming environment to execute Durable Functions efficiently. These frameworks are complete programming solutions that support advanced features (arbitrary composition, critical sections), but they are not well-suited for supporting large, complex analytics jobs. Because they involve manually combining operators into a DAG utilizing vendor LOCK-in API. For example, Netflix's Conductor~\cite{netflix_conductor}, Zeebe~\cite{zeebe}, and AWS Step Function~\cite{lambda_step_functions} use a JSON schema for authoring workflows, and Fission Workflows~\cite{fission}, Google Cloud Composer~\cite{google_cloud_composer}, and Fn Flow~\cite{fnflow}, are somewhat more code-based, as the schema is constructed in code. Without query optimizer, the customized jobs are error-prone and suboptimal, resulting in significant performance loss, and are seldom reused for streaming workloads. Instead, \flock supports Dataframe and SQL API to make streaming computation more accessible to users.

Data passing is a key challenge for chained cloud functions. everal systems, such as Pocket~\cite{klimovic2018pocket}, Locus~\cite{pu2019shuffling}, Caerus~\cite{zhang2021caerus}, and Cloudburst~\cite{sreekanti2020cloudburst}, have proposed solutions to improve the performance and cost-efficiency of ephemeral data sharing in serverless jobs using multi-tier remote storage, caching, or adding statefulness to serverless workflows.  Lambada~\cite{muller2020lambada} and Starling~\cite{perron2020starling} use S3 as exchange operators for shuffling large amounts of data, while SONIC~\cite{mahgoub2021sonic} employs a hybrid and dynamic approach to automatically choose data passing methods between serverless functions. In contrast to state-of-the-practice systems, \flock is the first system to build a streaming query engine for data passing on cloud function services using the payload of function invocations. This general solution is aimed at major cloud vendors without relying on external communication mediums.

\vspace{-2mm}
\section{Conclusion}

\flock is a step forward in real-time data analytics on FaaS platforms. The ability to leverage the on-demand elasticity of FaaS and the use of payload invocations for data passing, provides a new approach to stream processing that is cost-effective, low-latency and scalable. The elimination of external storage services makes \flock more efficient and easier to use than traditional systems.  As FaaS platforms continue to evolve and gain more widespread adoption, we expect to see more organizations embracing the use of \flock and similar systems to perform real-time data analytics.


\clearpage
\balance

\bibliographystyle{ACM-Reference-Format}
\bibliography{sample}

\end{document}